\documentstyle[epsfig,12pt]{article}
\newcommand{\be}{\begin{eqnarray}}
\newcommand{\ee}{\end{eqnarray}}
\newcommand{\ba}{\begin{array}}
\newcommand{\ea}{\end{array}}
\newcommand{\ur}[1]{(\ref{#1})}

\newcommand{\eq}[1]{eq.~(\ref{#1})}

\newcommand{\Eq}[1]{Eq.~(\ref{#1})}

  \newcommand{\la}[1]{\label{#1}}
  
  \def\beq{\begin{equation}}
  \def\eeq{\end{equation}}
  \def\beqr{\begin{eqnarray}}
  \def\eeqr{\end{eqnarray}}

\begin{document}
\thispagestyle{empty}

\vskip 5true  cm

\begin{center}
{\Large\bf To what distances do we know the confining potential?}
\vskip 2true cm

{\large\bf
Dmitri Diakonov$^{\diamond *}$ and Victor Petrov$^*$}
\\
\vskip 1true cm
\noindent
{\it
$^\diamond $NORDITA, Blegdamsvej 17, 2100 Copenhagen \O, Denmark \\
\vskip .2true cm
$^*$Petersburg Nuclear Physics Institute, Gatchina,
St.Petersburg 188 350, Russia}
\end{center}

\vskip 2.5true cm
\begin{abstract}
\noindent
We argue that asymptotically linear static potential is built in into
the common procedure of extracting it from lattice Wilson loop
measurements. To illustrate the point, we extract the potential by the
standard lattice method in a model vacuum made of instantons.  A
beautiful infinitely rising linear potential is obtained in the case
where the true potential is actually flattening. We argue that the flux
tube formation might be also an artifact of the lattice procedure and
not necessarily a measured physical effect.

We conclude that at present the rising potential is known for sure up
to no more than about $0.7$ fm. It may explain why no screening has
been clearly observed so far for adjoint sources and for fundamental
sources but with dynamical fermions.

Finally, we speculate on how confinement could be achieved even despite
the absense of the infinitely rising potential in the pure glue
theory.

\end{abstract}
\newpage
\section{Motivation}

In the last two decades it became a common place that confinement is
due to a linear rising potential between static probe quarks in the
4-dimensional pure Yang-Mills theory. Being a simple consequence of the
strong coupling expansion, an infinitely rising linear potential
becomes highly non-trivial in the weak coupling continuum limit.
Moreover, it contradicts all previous experience in physics with forces
decreasing with distances. Therefore, if proven correct, the linear
potential would be a most important discovery.

Unfortunately, a proof of the linear potential in the 4-dimensional
pure Yang--Mills theory is missing. Therefore, at present the only
source of knowledge about the behaviour of the static potential in the
pure glue world are lattice studies. However, lattice measurements have
statistical and, in many cases, systematic uncertainties due to
finite lattice spacing, volume and methodology used, and thus can never
substitute a rigorous mathematical proof of the linear potential. Being
a numerical experiment, lattice studies should be addressed by the same
questions as real-world experiments. In this case the main questions
are:
\begin{itemize}
\item To what distances the potential is reliably measured
\item To what accuracy it is measured; what are the systematic
uncertainties
\item What is the best-fit form of the potential in the range it is
reliably established
\end{itemize}

A clear answer to these questions is important for theoretical
models of confinement and for phenomenological applications.

An additional motivation for a critical analysis of what is presently
known about the static potential comes from the lattice studies
themselves. There have been measurements of the potential
between static sources belonging to the adjoint representation
\cite{M1,M2}, and also between sources in the fundamental
representation but with light dynamical fermions \cite{G,A}. In both
cases, contrary to the case of fundamental sources in the pure
glue theory, the rising potential is expected to flatten out at
certain separation owing to the screening or the `string-breaking'
effect. No clear evidence of flattening has been observed so far in
either of the cases.  Moreover, the potential between sextet sources in
the $SU(3)$ theory does not follow the triplet slope, as expected
\cite{M2}.

The non-observation of screening in cases where it is expected
is usually ascribed to a poor overlap of the quark creation
and annihilation operators, as given by the Wilson loop, with the
ground state at large separation between the probe sources. To override
this difficulty, it has been suggested to consider mixing of the
Wilson loop with other operators which do saturate at large
separations between the sources \cite{PW1,KS,S,PW2}. No wonder
that when one allows for a mixing with such operators the
diagonalization will always end up with the lowest eigenvalue
flattening at large separations. However, an essential finding of these
works is that when the separation between sources becomes large,
the Wilson loop effectively decouples from the lowest-energy state
characterized by a flattening potential.

What is the physical reason for a miniscule overlap of the
adjoint Wilson loop with the ground state of two widely separated
sources? How do we know that in the case of fundamental charges
the Wilson loop has, on the contrary, a sizeable overlap with the
ground state?  Unfortunately, the assumption that it is sizeable is
important to be able to extract the potential at large separations.

In all cases investigated until now in zero-temperature
4-dimensional theories the potential extracted from measuring Wilson
loops is compatible with a linear rise, both when it is expected or
unexpected.  This fact alone calls for a critical analysis of how the
potential is commonly extracted from Wilson loops.

The point of view which we advocate in this paper is that the
infinitely rising linear potential is built in by construction in the
commonly used procedure, and hence the systematic uncertainty has been
underestimated. In fact, the procedure is such that it is sufficient to
have the potential to be approximately linear in a limited range of
separations (around 0.5 fm) to get it infinitely rising at all
distances: the larger $r$, the better linear it comes out.

The conclusion is that if we do not assume from the start that the
potential {\em ought to be} linear (let us not forget: it is not
proved yet), and we do not assume that the overlap of the Wilson
loop with the ground-state potential is sizeable (and it is not
sizeable at least in two cases about which we know), the only
sure statement about the potential is that it is
aproximately linear up to about 0.7 fm, and still continues to rise.
Extraction of the potential above this scale, unfortunately,
involves assumptions.

\section{Standard procedure of extracting static potential \\
from Wilson loops}

Let us denote $W(r,t)$ a rectangular $r\times t$ Wilson loop averaged
over many gauge configurations. The standard transfer-matrix logic
says that it can be decomposed as a sum over intermediate states formed
by a quark-antiquark pair at separation $r$:

\beq
W(r,t)=\sum_n|C_n(r)|^2\exp\left[-V_n(r)t\right]
\la{decomp}\eeq
where $V_n(r)$ are the `potentials' for intermediate states $n$ and
$C_n(r)$ are the overlaps of these states with the concrete quark pair
creation operator.

To get ground-state potential $V(r)=V_0(r)$
one has to take the limit of large $t$. To be more quantitative, the
ground state is cut out from the sum \ur{decomp} at
$t\gg 1/\Delta E$ where $\Delta E$ is the energy splitting between the
ground and the next excited state. For a string of length $r$ this
splitting is expected to be
$\Delta E = V_1(r)-V_0(r)\sim 1/r$~\footnote{Recent measurements
\cite{JKM} indicate that certain excitations may be split even less
than by the expected $\Delta E = \pi/r$.}. Hence, in order to extract
the static potential one has to take Wilson loops with $t\gg r$.

Unfortunately, this key requirement can hardly be achieved for
physically interesting separations $r\geq 1$ fm. Let us imagine that
we want to measure the potential at a moderate separation of $r=1$ fm.
The $t$ side should be much much longer than 1 fm. We take a liberal
view and announce that $2\gg 1$, so let us take $t=2$ fm. The area of
the Wilson loop is then $2\;{\rm fm}^2$. The expected string tension is
$\sigma \simeq (430\;{\rm MeV})^2 \simeq 4.75\;{\rm fm}^{-2}$. The
expected value of the Wilson loop is then
$W\simeq \exp(-4.75\cdot 2) \simeq 10^{-4}$. To what accuracy do we
want to measure $W$? Let us take a moderate accuracy of 10\%, that is
we require $\Delta W\simeq 10^{-5}$. Since individual measurements of
$W$ fluctuate wildly in the range from -1 to 1, and the statistical
error $\Delta W$ goes as one over square root of the number of
independent measurements, it means that one needs an order of $10^{10}$
measurements. On a large lattice one can probably allow as much as
$10^4$ measurements of Wilson loops lying in different planes per one
gluon configuration, assuming they are statistically independent
\footnote{This is an optimistic estimate based on the correlation
length of about 1 fm. However, the Wilson loop is a peculiar object
for which the correlation length is infinity if one implies linear
confinement.}.

Summarizing this arithmetical exercise, we see that in order to
honestly measure the potential at a moderate 1 fm separation with a
modest 10\% accuracy one needs at least  $10^6$ statistically
independent gluon configurations!  This is beyond any
computer capacity either now or in near future: the typical number
of configurations used at present is no more than a few thousand.
With such statistics one can measure loops of areas no more than
$\simeq 1\;{\rm fm}^2$, even using the aforementioned liberal
assumptions.  With this murderous arithmetic one can wonder how any
quantitative statements can be made about the static potential at
separations beyond 0.7 fm.

To circumvent this difficulty, a link-smearing procedure has been
suggested \cite{APE,BS} presently used in most lattice studies.
The idea is to replace links along the spatial sides of the Wilson
loops by links smeared in other spatial directions, or by `fat' links.
Through this procedure the average of the Wilson loop increases many
times; it is ascribed to the larger overlap $|C_0(r)|$ of fat link
operator with the ground state, see \eq{decomp}. We shall see, however,
that this increase can be interpreted in another way.

If one is sure that by choosing an appropriate $\bar Q Q$ creation
operator one selects only the {\em ground} state contribution to the
decomposition \ur{decomp}, all what one needs is to check that
$W(r,t)$ follows a simple one-exponent decay with $t$. This is
performed not at $t\gg r$ but rather on the contrary at $t\ll r$.
The unfortunate `rule of a thumb' is that the area cannot exceed
$1\;{\rm fm}^2$ (because of the stringent statistics requirements),
therefore if one wants to measure the potential at $r=2$ fm, the $t$
side cannot exceed 0.5 fm, so that the exponential behaviour of
$W(r,t)$ in $t$ can be actually checked only up to quite small values
of $t$. The hope is that, once established at very low $t$, the same
time exponent will prevail at any $t$, therefore measurements at low
$t$ can give accurate values for the ground-state potential $V_0(r)$.

Examples of the plateaus in the quantity
$-\partial\ln W(r,t)/\partial t$ as function of $t$ for several values
of $r$ with $t\ll r$, is given in Fig.1. We show there the data
obtained by the Wuppertal group \cite{W} as being probably still a
record study. Using the $SU(2)$ gauge theory on lattices of volume up
to $48^3\times 64$ and $\beta$ up to 2.74, Bali, Schlichter and
Schilling were able to claim a linearly rising potential up to
distances up to 2.3 fm. In addition, a string formation over
physical distances up to 2 fm has been reported in this remarkable
study.

\begin{figure}
\centerline{\epsfxsize7.5cm\epsffile{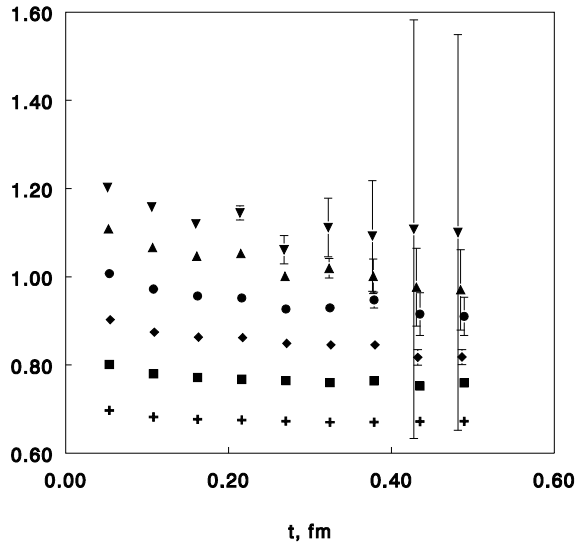}}
\caption[]{\it Effective potential
$V_{eff}(r,t)=\ln\left[W(r,t)/W(r,t+a)\right]$ as function of $t$ at
different values of $r$ from ref.\cite{W}. The data points correspond
(from bottom to top) to $r=0.65,\;0.98,\;1.30,\;1.62,\;1.95\;{\rm and}\;
2.25\;{\rm fm}$. The inverse coupling used $\beta=2.635$ corresponds
to the lattice spacing $a=0.0541\;{\rm fm}$. The data points have
been slightly scattered in the horizontal direction to resolve
the error bars. Courtesy G.Bali.}
\end{figure}

One can see from Fig.1 that the quality of the plateaus at
$r\leq 1\;{\rm fm}$ is quite good though there is trend
of the data to slope down as $t$ increases. At larger $r$
this trend becomes more pronounced, until the error bars explode
so that one can hardly get to any conclusions about the plateaus.
The procedure of extracting the potential is detailed in ref.\cite{W}
but basically it follows from the data in Fig.1. Having no objections
to the measurements {\em per se}, we still have doubts in their
interpretation, which we share below.

\section{The danger of misinterpretation}

We shall now explain why using the $t\ll r$ data is dangerous and may
result in large systematic errors in determining the potential, despite
the use of the smeared or fat spatial links.

First, it is known from numerical experience that smearing spatial
links leads to an essential increase of time-independent prefactor
in the average Wilson loops, but it is not clear why that necessarily
means an increase of the overlap with precisely the {\em ground} state
and not with some higher state or, even more probably, with some
complicated superposition of various states.

Second, even with the fat links one never gets the prefactor to be
exactly unity, hence one inevitably measures a contamination with
many states $V_n(r)$. One can, theoretically, imagine a case where
the true ground state has a tiny overlap with the smeared Wilson
loops: then it is next to hopeless to extract it. That such a case is
not totally academic is exemplified by the fact that the above
procedure, being applied to Wilson loops in the adjoint representation
or to the fundamental representation but with dynamical fermions, does
not show clear signals of the expected screening of the ground-state
potential \cite{M2,A}.

Third, most dangerous of all, when taking $t\ll r$, it can be
questioned why isn't it possible to turn the head by
$90^{{\tiny\rm o}}$ and call the long side ``$t$'' (instead of $r$)
and the short side ``$r$'' (instead of $t$). Then the exponential
falloff of the Wilson loop with the long side length $r$ (i.e. the
linear potential) is automatically guaranteed, because now it is time,
and the time is large. Taking fat links along the longer side only,
though it formally destroys the Euclidean $t\leftrightarrow r$
symmetry, does not override the danger of misinterpretation.

The interpretation of the Wilson loop with $t\ll r$ would be the
following. One creates a $\bar Q Q$ pair by the short (unsmeared)
side of the rectangular and separates them to the distance $t$.
By using a superposition of smeared links along the long side
one actually consideres not absolutely static $\bar Q$ and $Q$
but rather oscillating around fixed positions. Since smearing
is a kind of random walk in the transverse directions \cite{APE,BS},
the actual positions of $\bar Q$ and $Q$ can be said to be
Gaussian-distributed, with a width depending on the details of the
link-smearing procedure. Since Gaussian distribution is a wave function
for an oscillator potential, one can say that, by ways of smearing,
one studies the $\bar Q Q$ system in the normal gluon vacuum but with
quarks put in superficial oscillator potential wells centered at fixed
positions. Else, one can say that quarks are not infinitely heavy
anymore but experience normal zero-point oscillations in the gluon
vacuum. In either case one can introduce a hamiltonian effectively
describing the system. Its lowest-energy state depends on the $\bar Q Q$
separation (as given by the short side $t$) and is tested when one
measures the Wilson loop at large times, i.e. at large $r$.

Anyhow, for $r$ larger than $t$ and than the effective smearing
radius the average Wilson loop should behave as

\beq
W(r,t)\simeq \exp\left[-\tilde V_0(t)\cdot r\right],
\la{tvr}\eeq
where $\tilde V_0(t)$ is the ground-state energy of the $\bar Q Q$
pair oscillating somewhat around their separation $t$. If one now
determines the effective potential in the conventional way by
differetiating $\ln W(r,t)$ in respect to $t$, one finds

\beq
V_{eff}(r,t) =-\frac{\partial \ln W(r,t)}{\partial t}
\simeq \frac{d\tilde V_0(t)}{dt}\cdot r,
\la{effV}\eeq
i.e. a potential which is asymptotically linear in $r$. In order
to reproduce the plateau region in $t$ all one needs is to have
$\tilde V_0(t)$ approximately linear in $t$ in the limited
range of rather small values of $t$ where the `plateaus' are
actually checked, see Fig.1. Hence, from Fig.1 one finds that the
potential is approximately linear up to the endpoint of the
plateau region, that is up to 0.5~fm. The linear $r$ dependence of
the potential extracted from $V_{eff}$ at $t\ll r$ is then a
triviality: the larger $r$ is, the more exact \eq{tvr} becomes, the
better-quality `linear potential' one gets.

The `increase of the overlap with the ground state' assumed to be
achieved when one consideres smeared spatial links, gets a most
natural explanation from this point of view. The effective overlap is
defined as (see, e.g., \cite{W})

\beq
c_{eff}(r,t)=W(r,t)\,\exp\left[ V_{eff}(r,t)\cdot t\right]
\simeq \exp\left[-r\left(\tilde V_0(t)-t\frac{d\tilde V_0}{dt}
\right)\right].
\la{effo}\eeq
If $\tilde V_0$ is approximately linear in $t$ in the `plateau' range
of observations, the linear term in the parenthesis cancels,
however the constant part of $\tilde V_0$ does not. The
constant part is the quark self-energy. In perturbation theory it
diverges linearly for static quarks (corresponding to the unsmeared
spatial links) but only logarithmically for fluctuating sources
(corresponding to smeared links). Therefore, the constant part of
$\tilde V_0$ is much smaller for smeared links. This seems to be the
real reason why the `overlap' defined by \eq{effo} is increased by one
to two orders of magnitude when one goes from unsmeared to smeared
spatial links -- a fact which is not easy to explain in the standard
logic.

To conclude, {\bf there is a danger that the linear rising
potential is built in by construction into the procedure} of
extracting it from Wilson loop measurements at $t<r$, whereas it is
exactly what is so demanding to prove.

\section{Instanton ensemble}

We decided to check to what extent does the standard procedure work
by applying it to a model gluon vacuum for which the static potential
is known theoretically, namely to the random instanton
ensemble \footnote{The first comparison of Wilson loop measurements
with the instanton gas predictions has been made in ref. \cite{IMP}.}.
We take the simplest superposition ansatz of $N_+=N_-=N/2$ instantons
and antiinstantons ($I$'s and $\bar I$'s for short) in the singular
gauge,

\beq
A_\mu^a(x)=\sum_I^{N_+}A_\mu^{I\,a}(x)
+\sum_{\bar I}^{N_-}A_\mu^{\bar I\,a}(x),\;\;\;\;\;
A_\mu^{I\,a}(x)=\frac{O^{ai}\bar\eta^i_{\mu\nu}(x-z)_\nu 2\rho^2}
{(x-z)^2\left[(x-z)^2+\rho^2\right]}.
\la{sum}\eeq
(for $\bar I$'s the 't Hooft symbol $\bar\eta$ is replaced by $\eta$).
The $SO(3)$ orientation matrices $O^{ai}$ are taken to be random, as
well as the centers $z_\mu$. The sizes of $I$'s and $\bar I$'s are
distributed according to the probability

\beq
P(\rho)=\int_0^\rho d\rho\,\nu(\rho)=\frac{(\rho/\rho_1)^{b-4}}
{\left[1+(\rho/\rho_1)^{\nu-1}\right]^{\frac{b-4}{\nu-1}}},\;\;\;\;\;
b=22/3,\;\;\;\;\;0\leq P(\rho)\leq 1.
\la{sizedistr}\eeq
This distribution function follows 't Hooft's $\nu(\rho)\sim \rho^{b-5}$
regime at small sizes and falls off as $\nu(\rho)\sim 1/\rho^\nu$ at
large sizes. The parameter $\rho_1$ is related to the maximum of the
distribution at $\rho_0=\rho_1[(b-5)/\nu]^{1/(\nu-1)}$. We choose the
`conformal' $\nu=5$ power in numerics.

We have computed the averages of Wilson loops with various $r,t$ in
this random instanton ensemble. We have used $N_++N_-=128+128$
and $256+256$ $I$'s and $\bar I$'s put in a 4-dim cubic box of volume $V$.
The number of instanton configurations over which averaging was
performed varied from 800 for small loops to 1600 for larger ones.
The ratio of the most probable size $\rho_0$ to the average separation
between pseudoparticles $\bar R=(N/V)^{-1/4}$ was fixed to be
$\rho_0/\bar R=0.4$. With this ratio fixed, the measured potential
appears to be proportional, within errors, to the density $N/V$ which,
therefore, sets the scale both for the potential and for the units in
which the distances $r$ and $t$ are measured. To be specific, we choose
$\bar R =0.645\;{\rm fm}$, so that $\rho_0=0.258\;{\rm fm}$.

These values are compatible with the characteristics of the instanton
ensemble obtained from smearing the vacuum gluon configurations by the
RG mapping method \cite{DHK} though these authors find
the ensemble to be more dilute. However, at the moment we
are concerned not by the accurate description of the instanton ensemble
but by the methodological problem of extracting the static potential
from the Wilson loop measurements. Sufficient to say that a portion of
closely situated $I$'s and $\bar I$'s may be lost by the smearing
procedure, so that the above choice of parameters is not totally
unrealistic.

To ensure statistical independence of individual measurements
we made only one measurement per configuration of the Wilson loop
placed in the ($zt$) plane in the middle of an open box of length
2.62 or 3.11 fm (for the chosen instanton density). The
path-ordered exponents along rectangular loops were computed by solving
differential equations, or by taking products of `links' introduced
by hand to mimic the lattice procedure. With fields given by a
continuum formula, the first method is faster than the latter (for
given accuracy) since any standard routine of solving differential
equations makes the discretization in a more clever way than just
taking equal spacing independent of the field. We have found that one
needs `lattice spacing' not less than 0.06 fm to reproduce Wilson loops
for a typical configuration to an accuracy better than 5\%.

Choosing the lattice spacing to be 1/10 of the average separation
between instantons, i.e. 0.065 fm, we have performed a standard
link smearing procedure for the spatial sides of the loop, replacing
each link by a U-shaped `staple' lying in the transverse spatial
directions,

\beq
U_\mu(n)\rightarrow \frac{U_\mu(n)+\alpha \sum_{\nu=\pm x, \pm y}
U_\nu(n)U_\mu(n+\nu)U^\dagger_\nu(n+\mu)}{{\rm norm}},\;\;\;\;\;
\mu=z,
\la{staple}\eeq
with a variable weight $\alpha$.

We have found that this smearing has no effect, within statistical
errors, on the average of the Wilson loops, for any weight $\alpha$
varying between 0 and 1. The reason is that the instanton ensemble is
already smooth enough, so that no smearing of links is needed, unless a
small-size instanton happens to get inside a staple, which is
statistically a negligible effect. This should be contrasted with real
lattice calculations which are overwhelmed by ultraviolet noise, so
that smearing links has a dramatic effect. In any case our `overlaps'
$|C(r)|$ were not small, with fat links or without them.
According to the common reasoning, all what one needs then is to check
that $W(r,t)$ falls exponentially with $t$, even though $t$ is not much
larger than $r$.

We show the data for $W(r,t)$ in Fig.2 for $t$ ranging from 0.13 to
0.52 fm and $r$ ranging from 0.52 to 1.81 fm. For each value of $r$
the quantity $-\ln[W(r,t)]$ {\bf can be well fitted by a linear
dependence on $t$ even though $t$ is less than $r$}. There are no signs
that the curves wish to level off as $t$ increases. These are the
famous `plateaus' for the quantity $-\partial \ln(W(r,t) / \partial t$
which persuade optimists that the asymptotics in $t$ is
already reached, and that one can read off the static potential $V(r)$
as the slopes of the straight lines in $t$, and the overlap $|C(r)|^2$
as their intercepts.

\begin{figure}
\centerline{\epsfxsize8.0cm\epsffile{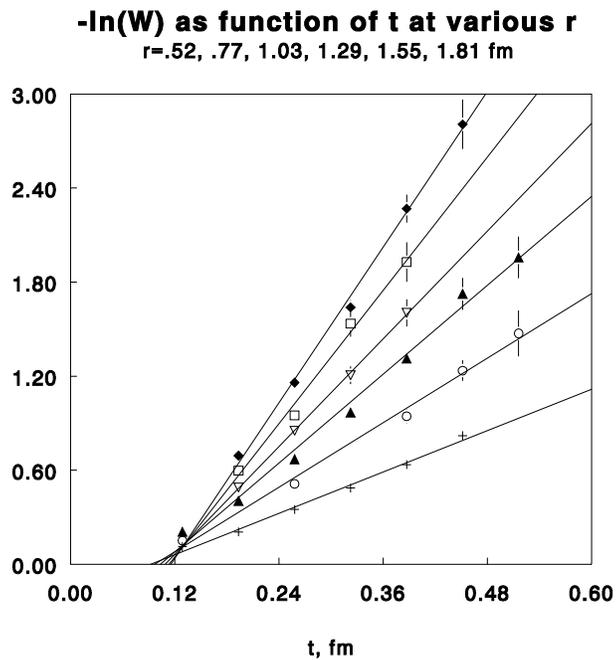}}
\caption[]{\it Seems to demonstrate that Wilson loops can be well
fitted by simple time exponents even at small values of $t$. }
\end{figure}

\begin{figure}
\centerline{\epsfxsize8.0cm\epsffile{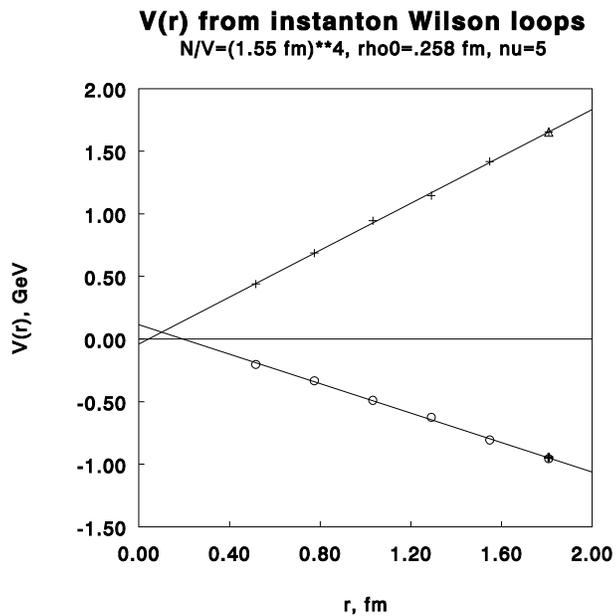}}
\caption[]{\it Effective potential $V_{eff}(r)$ (crosses) and `overlap'
$-\ln|C(r)|^2$ (circles) extracted from the data in Fig.2.
For the largest $r=1.81\;{\rm fm}$ the values of $V_{eff}(r)$ and
of $-\ln|C(r)|^2$ obtained from fat links are also shown for
illustration: the results coincide within errors with those
obtained from unsmeared links.}
\end{figure}

Following this common practice we plot $V(r)$ and $-\ln |C(r)|^2$
in Fig.3. A linear fit to $V(r)$ is quite impressive; it gives
the value of the `string tension' $\sigma \simeq (430\;{\rm MeV})^2$.
Naturally, there is no Coulomb $1/r$ term at small $r$, which emerges
from Gaussian quantum fluctuations of gluon field about whatever
background.

The potential $V(r)$ resulting from these Wilson loop measurements
is plotted again in Fig. 4, together with the theoretically-known
heavy-quark potential induced by instantons, which we explain below.

\section{Heavy-quark potential induced by instantons}

The leading term (in the density of instantons) of the
instanton-induced potential was given in ref. \cite{CDGWZ} without a
derivation. A derivation was presented in ref. \cite{DPP}, which allows
generalization to higher orders in density as well as to potentials
induced by objects different from instantons. A further generalization
to arbitrary groups and the representations for probe quarks has been
derived in ref. \cite{DP} which we cite here.

Let probe quarks belong to the representation $R$ of a gauge group $G$
whose dimensions are $d(R)$ and $d(G)$, respectively. For example,
for a fundamental representation of the $SU(N_c)$ group $d(R)=N_c$ and
$d(G)=N_c^2-1$. Since instantons are essentially $SU(2)$ objects
one has first of all to decompose the given representation $R$ in
respect to its $SU(2)$ content. For example, if one takes probe quarks
from the $SU(3)$ octet, it has two $SU(2)$ doublets with $J=1/2$, one
triplet with $J=1$ and one singlet with $J=0$, so that $\sum_J(2J+1)=
d(R)=8$. A fundamental representation of any $SU(N_c)$ group has only
one doublet (J=1/2), all the rest `particles' are $SU(2)$ singlets.
The instanton-induced potential can be also decomposed in contributions
of the $SU(2)$ multiplets,

\beq
V(r)=4\pi\frac{N}{V}\int_0^\infty\!d\rho\;\nu(\rho)\frac{1}{d(R)}
\sum_{J\in R}(2J+1)F_J(x),\;\;\;\;\;\;x=\frac{r}{2\rho}.
\la{genV}\eeq

Here $N/V$ is the $I$'s and $\bar I$'s density, $\nu(\rho)$ is their
size distribution normalized to unity and $F_J(x)$ are dimensionless
functions depending on the quark separation $r$ measured in units
of $2\rho$, they depend on the spin $J$ of the $SU(2)$ multiplet
inside the given representation $R$. These functions are given by
integrals over dimensionless variables $y=|z|/\rho$ and $t$ where
$|z|$ is the distance of an instanton from the axis drawn in the middle
between the two sources, and $t$ is the cosine of the angle between
$\overrightarrow {r}$ and $\overrightarrow {z}$:

\[
F_J(x)=\!\int_0^\infty\!\!dy\:y^2\!\!\int_0^1\!\!dt
\left(1\!-\!\cos\phi_+\!\cdot\cos\phi_-\!
-\!\frac{y^2-x^2}{\sqrt{x^2+y^2+2xyt}\sqrt{x^2+y^2-2xyt}}\sin\phi_+
\!\cdot\sin\phi_-\right),
\]
\beq
\phi_\pm=2\pi\sqrt{\frac{J(J+1)}{3}}
\left(\frac{\sqrt{x^2+y^2\pm 2xyt}}{\sqrt{x^2+y^2\pm 2xyt+1}}-1\right).
\la{FJ}\eeq

The functions $F_J(x)$ behave as $\sim x^2$ at small $x$; at large
$x$ they tend to constants depending on $J$. If the third moment of the
instanton size distribution is convergent the potential $V(r)$ flattens
out asymptotically to twice the renormalization of the heavy quark mass
\cite{DPP}

\beq
\Delta M = 16\pi\cdot 0.552\cdot\frac{N}{VN_c}\overline{\rho^3}
\;\;\;\;\;\;({\rm for\;quarks\;in\;the\;fundamental\;representation}).
\la{deltam}\eeq
If the size distribution happens to fall off as $\nu(\rho)\sim
1/\rho^3$ at large $\rho$ one gets a linear infinitely rising
potential \cite{DP}. However, such a size distribution means that
large instantons inevitably overlap, and the sum ansatz \ur{sum}
is not reasonable.

\Eq{genV} is actually the first term in the expansion of the potential
in the instanton density $N/V$ \cite{DPP}. We have evaluated the next
term and checked that it is much smaller than the first one in the
range of parameters of interest. Taking the size distribution
\ur{sizedistr} with the same set of parameters as in the numerical
simulations we get the potential shown in Fig.4. If one wants to change
the average size $\rho_0$ one has to rescale the $r$ axis; the
instanton density $N/V$ is just a scale factor of the potential as a
whole.

The correct instanton-induced potential starts to rise quadratically
with the separation, then in a rather long interval it remains
approximately linear but asymptotically it approaches $2\Delta M
\simeq 2\cdot 1.37\;{\rm GeV}$, for the chosen instanton distribution.

\begin{figure}
\centerline{\epsfxsize8.0cm\epsffile{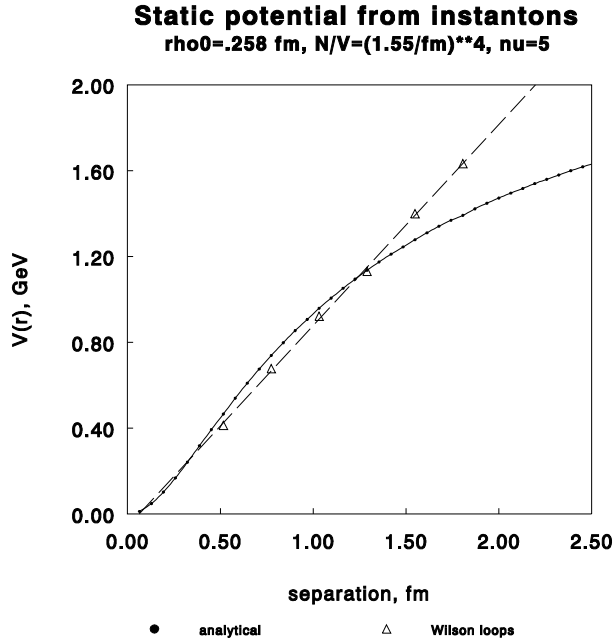}}
\caption[]{\it Static potential $V(r)$ as measured from Wilson loops
with $t<r$ (open triangles) versus theoretical prediction from the
instanton ensemble (solid line).}
\end{figure}

We see that the potential extracted by standard procedure from
Wilson loop measurements at $t<r$ reproduces the theoretical
expectation reasonably well at $r<1.3\;{\rm fm}$; at larger separations
the former continues to rise linearly while the latter flattens out.
\vskip .1true cm

This exercise illustrates that it might be dangerous to extract
the potential from Wilson loops with $t<r$ even though one observes
nice plateaus in $t$.

\section{Interpretation}

When one measures Wilson loops at $t\ll r$ the linear dependence of
$\log W(r,t)$ on $r$ is built in, because $r$ is the long side of the
rectangular; the larger $r$ is the better linear dependence in $r$ will
be seen by default. In order to get a `plateau' of
$\partial\log W(r,t)/\partial t$ in $t$ when $t$ is relatively small,
all one needs is the true potential to grow approximately linearly in a
limited range of separations corresponding to the `plateau' region of
$t$.  Such a behaviour is exemplified by a dense instanton ensemble,
as seen from Fig.4. Measurements with $t<r$ pick up the `string
tension' (i.e. the derivative $dV/dr$) from the steep part of the
potential at small to moderate separations, and continues it, by
construction, to arbitrarily large separations.

To discriminate between a hypothetical case of a first steeply
rising and then flattening potential, and a case of an infinitely
rising linear potential one has really to make measurements with
$t>r$ and not {\it vice versa}.

It has been shown recently in ref. \cite{Neg} that one can extract the
instanton-induced potential from Wilson loops only when one takes
$t\simeq (2\;{\rm to}\;3)\cdot r$, as it should be expected from
general considerations. In Fig.5 we plot the results of ref. \cite{Neg}
in comparison to \eq{genV}. We see, first, that the instanton-induced
potential is, to a good accuracy, proportional to the instanton density
$N/V$, which justifies the use of the first virial expansion term
\ur{genV}; second, that this formula reproduces well the potential
extracted from Wilson loops with $t\simeq (2\;{\rm to}\;3)\cdot r$.

It may be argued that the instanton model does not correspond to
any well-defined Hermitean hamiltonian, and hence one cannot,
generally speaking, write the spectral decomposition for Wilson loops,
\eq{decomp}. We find, however, that the average $W(r,t)$ is a positive
function monotonously decreasing with $t$ at fixed $r$ and with $r$ at
fixed $t$. The same has been observed in ref. \cite{Neg}.
Therefore, $W(r,t)$ can, in fact, be decomposed as in \eq{decomp} with
positive coefficients and with positive effective potential
$V_{eff}(r,t)$, see.  \eq{effV}. In that respect the instanton model
does not differ from the full Yang--Mills theory.

Summarizing, measurements of Wilson loops with $t\gg r$ do reliably
reproduce the true potential. This is not the case for loops with
$t<r$.

\begin{figure}
\centerline{\epsfxsize8.0cm\epsffile{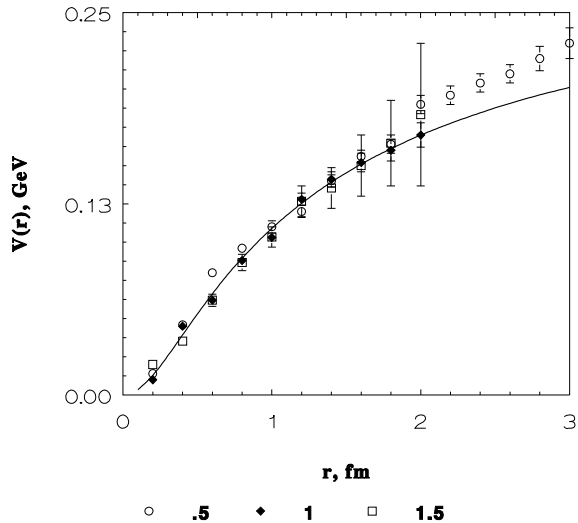}}
\caption[]{\it Potential extracted from Wilson loops at $\;t\gg r$
{\rm \cite{Neg}} in comparison with the theoretical curve \ur{genV}.
The data at $N/V = 0.5,\;1.0$, and $1.5\;1/{\rm fm}^4$ are divided
by factors $0.5,\;1.0$ and $1.5$ to demonstrate scaling in $N/V$.
The theoretical curve is computed with instanton size distribution used
in ref. {\rm\cite{Neg}}: it is different from the one used in
Figs.2-4.}
\end{figure}

A very important phenomenon whose observation would strongly support
the string picture in general and the infinitely rising linear
potential in particular, is a formation of a flux tube as the
separation between source quarks increases. This phenomenon has been
also studied in much detail and with unprecedented precision in ref.
\cite{W} by ways of measuring correlations of Wilson loops with
plaquettes placed inside the loops \footnote{To our knowledge, such
measurements were first performed in refs. \cite{H,D}.}.

As in the case of the potential, an irreproachable way to extract the
fields created by a pair of static quarks would be to use Wilson loops
with $t\gg r$. Unfortunately, in this case the statistics requirements
are even more disastrous than in the case of extracting $V(r)$, since
the signal-to-noise is smaller than for the Wilson loop itself.
Therefore, the authors of ref. \cite{W} are forced to consider the
opposite limit, $t\ll r$, again assuming that the ground state is cut
out by using fat spatial links.  In fact, the side called $t$ has been
taken below 0.5 fm while the long side, called $r$, was taken up to
more than 2 fm. As the long side increases, the authors observe a
certain flattening of the fields extent in the transverse plane, which
is interpreted as a flux tube formation.

Unfortunately, this interpretation suffers from the same ambiguity
as the extraction of the potential. With $t\ll r$ one can view
the fields as created by quarks oscillating about the moderate
separation $t$ (oscillations are the result of the link-smearing
procedure) but existing during a long time $r$. It is then a
{\bf triviality} that at large $r$ the average fields become constant
in time, i.e. in $r$. To check that there is a real string formation
and not a misinterpretation one has to make sure that the flux tube is
not thinning away as $t$ gets much larger than $r$. This seems to be a
formidable task, in view of the arithmetic presented in section 2
\footnote{In ref. \cite{W} stability of the fields between sources
as function of $t$ is illustrated in Figs. 17-19 where a rather
moderate separation $r=0.5$~fm has been used, and with $t$ varying
also up to $0.5$~fm. Despite moderate parameters of the loop the
error bars exceed 50\% at the largest value of $t$ shown there, i.e.
0.5~fm.}.

\section{Discussion}

There is a paradox in lattice measurements mentioned in the first
section: at zero temperatures no screening of the rising potential has
been clearly observed so far in situations where screening is expected.
This is the case of a pure glue theory with adjoint sources
\cite{M1,M2} and the case of fundamental sources but with dynamical
fermions \cite{G,A}. This lack of screening is deduced from measuring
Wilson loops with `fat' links at $t\ll r$ since it is statistically
impossible to study the opposite limit.  An obvious resolution of this
paradox would be that in neither of the cases one measures the true
potential but rather an automatic continuation of the potential from
the region with maximal $dV/dr$ to larger separations. It should be
mentioned that quite recently a clear indication of screening
has been observed with dynamical fermions in $d=3$ in ref. \cite{T}
where rather large values of $t$ for the Wilson loop were shown to be
needed, however, at the cost of using coarse lattices.

Would a saturation of the static potential for fundamental sources in
pure glue theory at some finite value of $V_\infty$ mean that there is
no confinement?  Not necessarily.  Quarks are confined in full QCD
despite there is no long distance force, and, by the way, so are gluons
despite the adjoint sources should be also screened. One has to
understand why it is so.

The real world has quarks, both light and heavy ones. The physics of
light quarks is strongly dominated by the effect of spontaneous chiral
symmetry breaking. It results in light quarks acquiring a dynamical
(or constituent) mass of about $M_{const}\simeq 350\;{\rm MeV}$ with
Goldstone pions becoming the lowest excitations in the spectrum.
Therefore, light quarks might not exist as asymptotic states: instead of
producing a light quark-antiquark pair it is energetically favourable
to produce one or several pions. Mathematically, it would correspond to
the quark propagator with momentum-dependent mass, having singularities
only on the `second Riemann sheet' under the cut starting from the
pion threshold.

As to heavy quarks, if $\Delta M=V_\infty/2$ happens to be larger
than approximately $M_{const}$ the heavy quarks would be unstable
under a decay to $B$ or $D$ mesons. The case is to some extent similar
to electrodynamics with charges $Z>137$: such particles are unstable
under a production of $e^+e^-$ pairs and therefore cannot exist as
asymptotic states. The heavy quarks might be thus confined too
\footnote{The ``$Z>137$'' scenario of confinement has been advocated
for many years by V.N.Gribov \cite{Gr1,Gr2}.}.

For the instanton vacuum it is possible to quantify the condition that
the renormalization of heavy quark mass is larger than the light
constituent quark mass. The quantity $\Delta M$ has been given above
in \eq{deltam} while the constituent quark mass is given by the
equation \cite{DP2}

\beq
M_{const}\simeq 1.45\cdot
2\pi\sqrt{\frac{N}{VN_c}}\sqrt{\overline{\rho^2}}.
\la{constm}\eeq
Notice that $\Delta M$ is linear in the instanton density while
$M_{const}$ is proportional to its square root. This important
circumstance is due to the fact that nonzero $M_{const}$ is an order
parameter for chiral symmetry breaking. The condition that
$\Delta M > M_{const}$ reads

\beq
\frac{N}{VN_c}\frac{\left(\overline{\rho^3}\right)^2}{\overline{\rho^2}}
>0.1,
\la{dense}\eeq
meaning that the instanton medium should be sufficiently dense but not
necessarily very dense. With this condition fulfilled, there is a good
chance of getting confinement of quarks even in the case where the
static potential levels off.

Finally, we would like to draw attention to persistent warnings
by Grady that certain phenomena usually associated with confinement
in pure glue theory might be, in fact, relics of the strong coupling
regime. These include the density of abelian monopoles and center
vortices \cite{Grady1}, the formation of percolating monopole clusters
\cite{Grady2} and the value and the linearity of the static potential
itself \cite{Grady1,Grady3}.

\section{Conclusions}

The answer to the question put in the title depends, unfortunately,
on the standpoint of a person. A pessimist would say that we are still
in a kind of strong coupling regime in certain lattice measurements,
and that it is still too early to draw any conclusions about the
behaviour of the static potential.

An optimist would say that the potential is proven to be linear
up to an enormous 4~fm separation \cite{JKM}, however it implies that
the asymptotics of Wilson loops is reached at incredibly small
$t<0.25\;{\rm fm}$. We remind the reader that at present one cannot
measure Wilson loops with areas exceeding $\simeq 1\;{\rm fm}^2$
for statistical reasons.

We think that a more weighted conclusion which can be made from
lattice measurements is that the static potential is still rising
at distances about 0.5 -- 0.7~fm but its precise form is unknown
beyond that separation. Similarly, we would avoid making
definite conclusions on string formation.

The link-smearing procedure being quite useful for measuring point
correlation functions seems to be extremely dangerous for measuring
nonlocal quantities such as Wilson loops since there is a real risk
of getting the string and the linear potential by mere construction of
the procedure, if one restricts oneself to measurements at $t<<r$.
We have illustrated that by applying the procedure to the model gluon
vacuum made of instantons for which the potential is known theoretically.

A force that is not decreasing with the distance is a feature never
before encountered in 3+1 dimensional physics. If correct, this
statement is so important that it deserves to be demonstrated beyond
any reasonable doubt.

\vskip .5true cm
{\bf Acknowledgments}
\vskip .5true cm
We would like to thank the participants of the workshop "Instantons and
Monopoles in the QCD Vacuum" (Copenhagen, June 1998) where the issue
raised here has been discussed in detail, especially to Pierre
van Baal, Gunnar Bali, Jeff Greensite and Arjan van der Sijs. D.D. is
most grateful to Gunnar Bali for a correspondence and for providing
with certain data from ref. \cite{W}. A correspondence with
K.J.Juge, C.Legeland and J.Negele is also gratefully acknowledged.
This work was supported in part by grants INTAS-RFBR 95-0681 and
RFBR 9727-15L.

\end{document}